\documentclass[12pt,preprint]{aastex}

\usepackage{graphicx} 

\shorttitle{Non-detection of magnetic fields in the central stars of the planetary nebulae NGC 1360 and LSS 1362}
\shortauthors{F. Leone}

\begin{document}

\title{Non detection of magnetic fields in the central stars of the
planetary nebulae NGC1360 and LSS1362}

\author{Francesco Leone\altaffilmark{1},  Mar\'{i}a J. Mart\'{i}nez Gonz\'alez\altaffilmark{2,3},  Romano L.M. Corradi\altaffilmark{2,3}, Giovanni Privitera\altaffilmark{1},  Rafael Manso Sainz\altaffilmark{2,3}}

\affil{$^{1}$ Universit\`a di Catania, Dipartimento di Fisica e Astronomia, Sezione Astrofisica, Via S. Sofia 78, I--95123 Catania, Italy\\
$^{2}$ Instituto de Astrof\'isica de Canarias, E-38200 La Laguna, Tenerife, Spain\\
$^{3}$ Departamento de Astrof\'isica, Universidad de La Laguna, E-38205 La Laguna, Tenerife}

\begin{abstract}
The presence of magnetic fields is an attractive hypothesis for shaping PNe. We report on observations
of the central star of the two Planetary Nebulae NGC1360 and LSS1326. We performed spectroscopy on
circularly polarized light with the {\em FOcal Reducer and low dispersion Spectrograph} at the Very Large
Telescope of the European Southern Observatory. Contrary to previous reports 
(Jordan et al. 2005, A\& A, 432, 273), we find that the effective magnetic field, that is the 
average over the visible stellar disk of longitudinal components of the magnetic fields, is null 
within errors for both stars. We conclude that a direct evidence of magnetic fields on the central 
stars of PNe is still missing --- either the magnetic field is much weaker ($<$ 600 G) than 
previously reported, or more complex (thus leading to cancellations), or both. Certainly, 
indirect evidences (e.g., MASER emission) fully justify further efforts to point out the 
strength and morphology of such magnetic fields. 
\end{abstract}


\section{Introduction}
A large fraction of planetary nebulae (about 80\%) are bipolar or
elliptical rather than spherically symmetric. Many of them also
harbor complex structures on small scales, such as knots, filaments,
jets and jet-like features, etc. (see e.g. \cite{Corradi06}).  But the
reason for the departure during the PN phase from the spherical
symmetry that has characterized most of the evolution of the
progenitor stars, is still a matter of debate. Modern theories invoke
magnetic fields, among other causes, to explain the rich variety of
aspherical components observed in PNe (see the review by
\cite{BalickFrank2002}).  The presence of magnetic fields would indeed
help to explain some features of the complicated shapes of planetary
nebulae, as the ejected matter is trapped along magnetic field lines.
There are several ways magnetic fields can be created in the vicinity
of planetary nebulae. Magnetic fields can be produced by a stellar
dynamo during the phase when the nebula is ejected. It is also
possible that the magnetic fields are fossil relics of previous stages
of stellar evolution \citep{Blackman01}. Under most circumstances, the
matter in stars is so highly electrically conductive that magnetic
fields can survive for millions or billions of years. In both cases,
the magnetic field combined with other physical processes including
stellar rotation, winds interaction, interaction wit the interstellar
medium, and the dynamical action of evolving photoionization fronts,
would produce the complex morphologies observed in PNe.

Until recently, the idea that magnetic fields are an important
ingredient in the shaping of planetary nebulae was mostly a
theoretical claim, since no such magnetic field was measured in the
nebulae themselves. To obtain direct evidence for the presence of
magnetic fields in planetary nebulae one can focus on their central
stars, where the magnetic fields should have survived.

\cite{Jordan05} report the detection of magnetic fields in the central
star of two non-spherical planetary nebulae, namely NGC1360 and LSS
1362. The claim is based on circular light spectropolarimetry carried
out with the {\it FOcal Reducer and low dispersion Spectrograph}
(FORS) at the Very Large Telescope (VLT) of the European Southern
Observatory (ESO).  The impact of \cite{Jordan05} paper is testified
by the 50 citations counted at the end of 2010, and increasing efforts
to include magnetic fields in theory of planetary nebulae.  As an
example, \cite{Tsui08} performed MHD calculations to model an
equatorial plasma torus in around the central stars of PNe.

Because of the achieved noise level and adopted set-up,
\cite{Jordan05} could not obtain a direct measurement of these
magnetic fields (see their figures 2 and 3). Controversial values were
also obtained from different Balmer lines.  These authors found
necessary to perform a large number of simulations to associate a
statistical significance to their results. 

In this paper, we present the results of new spectropolarimetric
measurements of NGC1360 and LSS1362 obtained on Dec. 22, 2010 with
FORS2 at the VLT, at higher signal to noise ratio, reciprocal
dispersion and spectral resolution than previously done by Jordan and
co-workers, with the aim to finally obtain a direct evidence of magnetic fields
on the surface of the central star of these planetary nebulae.

\begin{figure*}
\begin{center}
\includegraphics[width=15cm,height=6.3cm]{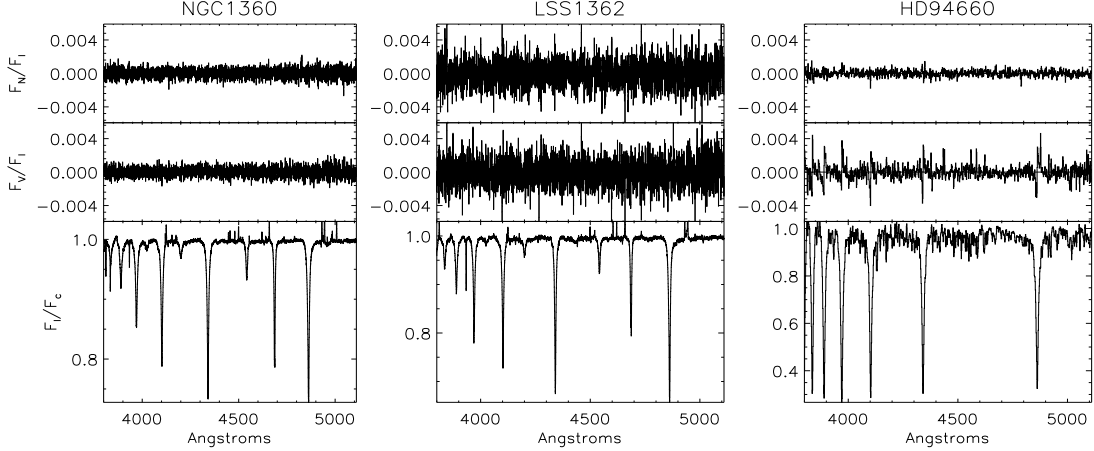}
\caption[]{Observed Stokes ${\cal F}_I/{\cal F}_c$, ${\cal F}_V/{\cal F}_I$ and ${\cal F}_N/{\cal F}_I$ (see text) of the
  central stars of the PNe NGC1360 and LSS1362. The Zeeman signature in the Stokes $V/I$   spectra of Balmer lines 
 is absent in the case of PNe and well visible in the magnetic star HD94660.}
\label{fig_observed}
\end{center}
\end{figure*}
\begin{figure*}

\begin{center}
\includegraphics[width=15cm,height=4.3cm]{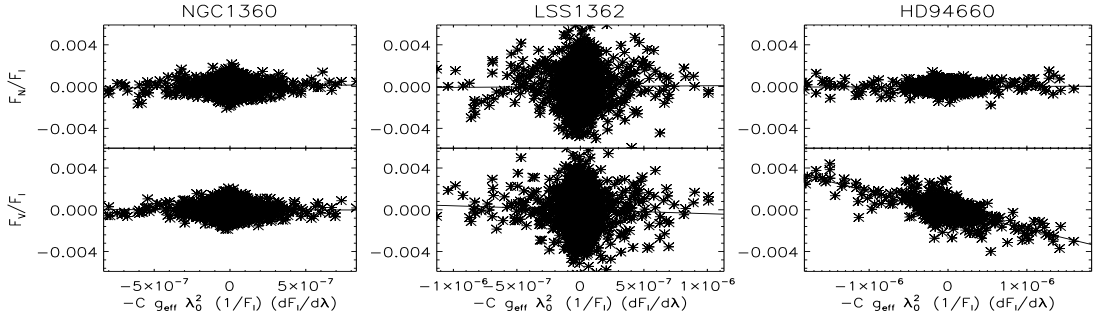}
\caption[]{The slope of Stokes ${\cal F}_V/{\cal F}_I$ vs. 
 $-C g_{\rm eff} \lambda_0^2 \frac{1}{{\cal F}_I} \frac{{\rm d}{\cal F}_I(\lambda)}{{\rm d}\lambda}$ measures the effective magnetic field (B$_e$). 
No slope is observed in the case of PNe. A B$_e$ value consistent with the
literature data is obtained for the magnetic star HD94660. The zero slope of ${\cal F}_N/{\cal F}_I$ spectra
shows no spurious polarization effects.}
\label{fig_slope}
\end{center}
\end{figure*}

\section{Observations and data reduction}

\cite{Jordan05} observed NGC1360 and LSS1362 with the FORS1
spectrograph at the VLT using the R600B+22 grating, a 0.8 arcsec slit, a MIT
24$\mu$m CCD and adopting a total exposure time of 624~s.  
Dispersion was 1.18\AA~pix$^{-1}$, and spectral resolution R$\sim$1200.
Circular spectropolarimetry was carried out with the standard
procedure consisting of a series of exposures at two different angles
of the $\lambda/4$ retarder with respect to the Wollaston axes, namely
$\alpha=+45$ and $-45^o$.

\begin{table*}
\tiny
\begin{center}
\label{Tab_results}
\caption{B$_{\rm eff}$ measurements from single Balmer lines, g$_{\rm eff} = 1$.
Average values are from a simultaneous fit of all available Balmer lines.}
\begin{tabular}{c|cc|cc|cc}\hline\hline
    & \multicolumn{2}{c|}{NGC 1360} & \multicolumn{2}{c|}{LSS 1362} & \multicolumn{2}{c}{HD 94660}\\
HJD/B/Seeing & \multicolumn{2}{c|}{2455553.503/10.99/1''} & \multicolumn{2}{c|}{2455553.832/12.27/2''} & \multicolumn{2}{c}{2454181.658/6.02/0.6''} \\
 & B$_{\rm eff}(V/I)$ [G] & B$_{\rm eff}(N/I)$ [G]  & B$_{\rm eff}(V/I)$ [G] & B$_{\rm eff}(N/I)$ [G]  & B$_{\rm eff}(V/I)$ [G] &  B$_{\rm eff}(N/I)$ [G]  \\
\rm All        & $+154\pm113$  & $+121\pm103$   & $-337\pm 286$      & $+34\pm221$      & $-1950\pm 71$      & $-14\pm29$  \\
\hline
H$_{\theta}$   & -- & --  & --  & --                                                        & $-1982 \pm 300$  & $-313\pm 122$  \\
H$_{\eta}$     & -- & --  & --  & --                                                        & $-2051 \pm 371$  & $+293\pm 151$  \\
H$_{\zeta}$    & $-78\pm 1656 $ & $-659\pm 1502$  & $+2094 \pm 3580$  & $+1506\pm 2870$  & $-2218 \pm 378$  & $+109\pm 154$  \\
H$_{\epsilon}$ & $+220\pm 867$ & $+200\pm 787$  & $+372 \pm 1613 $  & $-533\pm 1293  $  & $-1680 \pm 369$  & $-131\pm 150$   \\
H$_{\delta}$   & $-556\pm 523 $ & $-217\pm 474 $  & $-528\pm 1223   $  & $+1137\pm 980 $  & $-1805 \pm 364$  & $-181\pm 149$  \\
H$_{\gamma}$   & $+104\pm 379$ & $+35\pm 344 $  & $-843\pm 882    $  & $+389\pm 707  $  & $-1940 \pm 362$  & $+370\pm 147$  \\
H$_{\beta}$    & $+450\pm 318$ & $+325\pm 288$  & $-236\pm 734    $  & $-154\pm 588   $  & $-1973 \pm 351$  & $-88\pm 143$  \\\hline
\end{tabular}
\end{center}
\end{table*}

To improve the precision of the measurements, we observed NGC1360 and
LSS1362 for 3072~s each with the R1200B+97 grating, a 0.5 arcsec slit
and the E2V blue-optimized 15 $\mu$m CCD.  This setup results in a
linear dispersion equal to 0.35\AA~pix$^{-1}$ and R$\sim$2700 as
measured in the spectral lines of arcs.  The effects of linear
dispersion and line broadening in measuring stellar magnetic fields
are discussed in \cite{Leone00}.  To handle cosmic rays, observations
were split in a series of exposures of 256 seconds switching the
angle $\alpha$ between $+45$ and $-45^o$. On the coadded spectra of NGC1360 and LSS1362, we measured S/N $\sim$ 2400 and 800 respectively.

The combination of the {\it o}-rdinary and {\it e}-xtraordinary beams
emerging from the polarizer to measure the circular polarization
degree is critical.  There is a time independent (instrumental)
sensitivity $G$, for example due to a pixel-to-pixel efficiency,
together with a time dependent sensitivity $F$ of spectra obtained at
different $\alpha$ angles for example due to variation of sky
transparency and slit illumination. Photon noise dominated Stokes $I$
and $V$ can been obtained from the recorded spectra at
$\alpha$=$+45$$^o$ and $-45$$^o$:
\begin{eqnarray}
S_{+45^\circ,o} = & 0.5\,(I + V) G_{o} F_{+45^\circ} \nonumber  \\
S_{+45^\circ,e} = & 0.5\,(I - V) G_{e} F_{+45^\circ} \nonumber \\
S_{-45^\circ,o} = & 0.5\,(I - V) G_{o} F_{-45^\circ}\nonumber \\
S_{-45^\circ,e} = & 0.5\,(I + V) G_{e} F_{-45^\circ} \nonumber
\end{eqnarray}
Hence:
$$
\frac{V}{I} = \frac{R_V - 1}{R_V + 1}\hspace{1.5cm}{\rm with}\hspace{0.38cm}R_V^2 = \frac{S_{+45^o,o}/S_{+45^o,e}}{S_{-45^o,o}/S_{-45^o,e}}\nonumber
$$

We have reduced the data following the previous relations as in \cite{Leone07}.
In addition to $V/I$ we have also computed the {\it Noise} spectrum: 
$$
\frac{N}{I} = \frac{R_N - 1}{R_N + 1}\hspace{1.5cm}{\rm
with}\hspace{0.38cm}R_N^2 = \frac{S_{+45^o,o}/S_{-45^o,e}}{S_{-45^o,o}/S_{+45^o,e}}
$$
In an ideal polarimeter, signal extraction and wavelength calibration of ordinary and extraordinary spectra,
$N/I$ is null and its {\it absolute} error equal to $\rm (N_{total})^{-1/2}$, where 
$\rm N_{total}$ is the total number of photons. Any anomalous behavior of $N/I$ would be present, at the same level,
in Stokes $V/I$ by definition.

To test our capability to recover the circular polarized signal from
FORS spectra and measure stellar magnetic fields, we have applied the previous
procedures to the spectropolarimetric data, obtained from ESO archive, of the magnetic star
HD94660, whose field is at the intensity level of NGC1360 and LSS1362
as claimed by \cite{Jordan05}.  \cite{Landstreet00} have shown that the
magnetic field of HD94660 is variable with a 2700 day period between
$-$1.8 and $-$2 kG.  
Projected rotational velocities are also comparable, as NGC1360 shows
$v_e\sin\ i <$ 20 km\ s$^{-1}$ \citep{Garcia08} and HD94660 $<30$
km\ s$^{-1}$ \citep{Levato96}. We did not find any estimate in the
literature for the projected rotation velocity of LSS1362, whose
spectral lines appears in our spectra as broad as NGC1360 lines.

Fig. \ref{fig_observed} shows the observed spectra of NGC1360, LSS1362 and HD94660.

\section{Measuring magnetic fields}
High resolution circular spectropolarimetry of metal
lines gives the possibility to distinguish photospheric
regions with positive and negative magnetic fields, as for instance
done on HD24712 by Leone \& Catanzaro (2004, R =115 000).
It is also proved useful at moderate resolution (Leone
\& Catanzaro 2001, R = 15 000), but is still prohibitive to detect
magnetic fields of faint stars. For faint white dwarfs, Angel
\& Landstreet (1970) introduced a method based on narrowband
($\sim$30~\AA\ ) circular photopolarimetry on the wings of the
H$_\gamma$ Balmer line.

In the weak field approximation
for stellar atmospheres \citep{Landstreet82, Mathys89}, the
disk integrated Stokes-V parameter (the difference between the opposite circular polarized intensities)
${\cal F}_V$, is proportional to the derivative of the intensity flux ${\cal F}_I$:
\begin{equation}
\frac{{\cal F}_V}{{\cal F}_I}=-C g_{\rm eff} \lambda_0^2 \frac{1}{{\cal F}_I} \frac{{\rm d}{\cal F}_I(\lambda)}{{\rm d}\lambda} B_{\rm eff}
\end{equation}

where $C=-4.67\times 10^{-13}$~G$^{-1}$\AA$^{-1}$, $g_{\rm eff}$ is the effective Land\'e factor of the transition, $\lambda_0$ the wavelength in \AA, and

\begin{equation}
B_{\rm eff}=\frac{3}{2\pi}\int_0^{2\pi}{\rm d}\phi\int_0^1 B_\parallel \mu{\rm d}\mu
\end{equation}
is the longitudinal component of the magnetic field ($B_\parallel$) integrated over the stellar disk.

The slope of the linear regression of ${\cal F}_V/{\cal F}_I$ versus $-C g_{\rm eff} \lambda_0^2 \frac{1}{{\cal F}_I} \frac{{\rm d}{\cal F}_I(\lambda)}{{\rm d}\lambda}$ (forced to pass through the origin), gives the effective magnetic field.
In other words, we minimize the $\chi^2$ merit function

\begin{equation}
\chi^2=\sum_{ij}\frac{1}{\sigma^2}\left[({\cal F}_V/{\cal F}_I)^i_j+C\ g^i_{\rm eff}\  (\lambda_0^{i})^2 \
\frac{1}{({\cal F}_{I})^i_j} \frac{{\rm d}({\cal F}_{I})^i_j}{{\rm d}\lambda} {\rm{B_{\rm eff}}}\right]^2
\end{equation}

where the standard deviation of the noise $\sigma$ is independent of the spectral line $i$ and wavelength $j$.
If $({\cal F}_I')^i_j=(\lambda_0^i)^2\ g^i_{\rm eff}\ \frac{1}{({\cal F}_I)^i_j} \frac{{\rm d}({\cal F}_I)^i_j}{{\rm d}\lambda}$, after some algebra:
\begin{equation}
B_{\rm eff}=-\frac{\sum_{ij}({\cal F}_V/{\cal F}_I)^i_j({\cal F}_I')^i_j}{C\sum_{ij}[({\cal F}_I')^i_j]^2}
\end{equation}
while the error is obtained from the covariance matrix:
\begin{equation}
\delta B_{\rm eff}=\pm \sqrt{\Delta\chi^2}\frac{\sigma}{C\sqrt{\sum_{ij}[({\cal F}_I')^i_j]^2}}
\end{equation}
where $\Delta \chi^2$ are isocontours of the $\chi^2$ function that contain a certain confidence level. The values of $\Delta \chi^
2$ are tabulated and depend on the number of degrees of freedom. In our case, with only one degree of freedom, $\Delta \chi^2=1, 4,
 9$ for a confidence level of 68.3\%, 95.4\%, and 99.7\%, respectively.

\begin{figure*}
\begin{center}
\includegraphics[width=15cm,height=4.3cm]{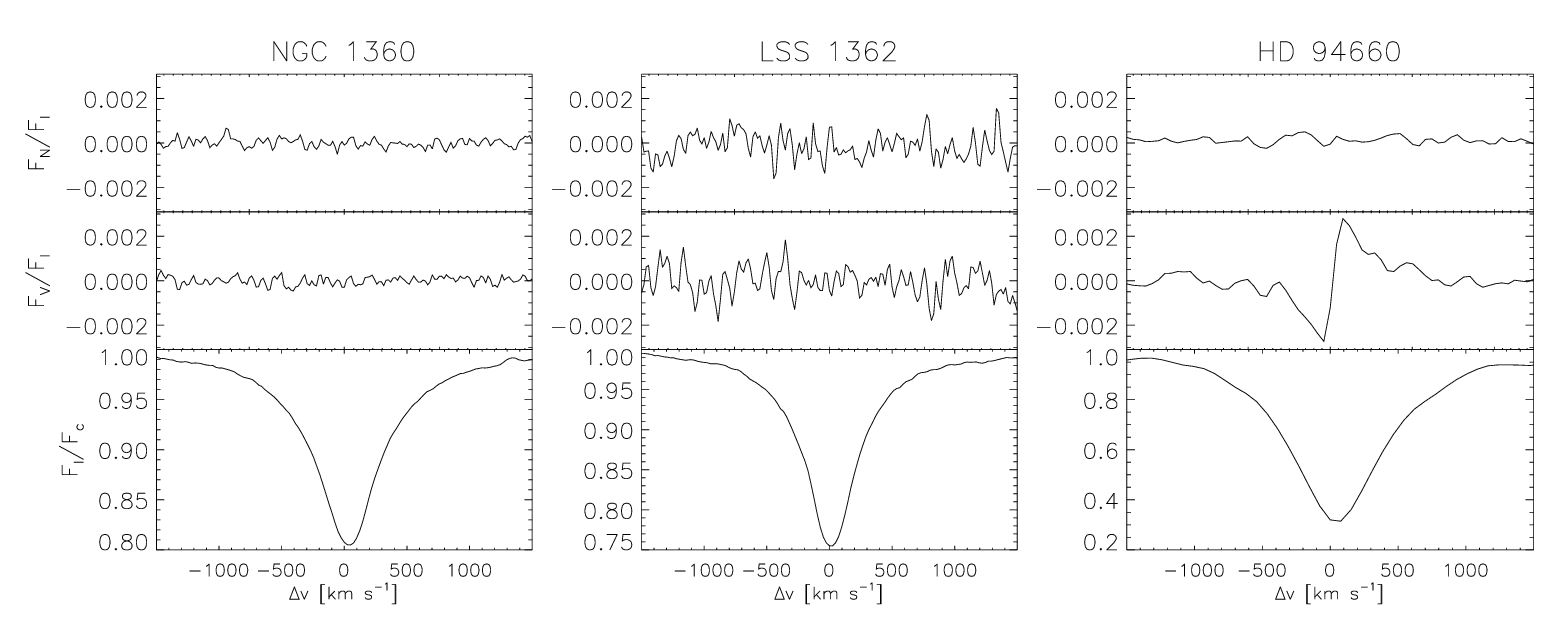}
\caption{Mean Stokes ${\cal F}_I/{\cal F}_c$,  ${\cal F}_V/{\cal F}_I$ and ${\cal F}_N/{\cal F}_I$ Balmer line profiles of the central stars of PNe NGC1360 and  LSS1362.
For comparison we report the same profiles of the magnetic star HD94660, whose effective magnetic field is $-1950$ G.}
\label{fig_folded}
\end{center}
\end{figure*}

Results of our measurements are listed in Table 1.
No single magnetic field fits the polarization of all Balmer
lines, as is instead observed in the reference magnetic star
HD94660. This is consistent with the absence of Zeeman
signal in any line for the two PN central stars. Taking into
account the information of all spectral lines consistently (i.e.,
fitting all the signals with the same field), the magnetic fields
obtained are $154\pm 113$ and $−337 \pm 286$~G, for NGC1360 and LSS 1362, respectively. In other words,
the magnetic field is essentially undetermined within errors,
with the most probable value compatible with the observations
lying well below the kG. It is important to stress that
we measured the magnetic field by minimizing the sum of the
merit function ($\chi^2$) for all spectral lines simultaneously, and
not as the weighted average of the magnetic fields obtained
from individual spectral lines, which is not correct.

Magnetic fields $\sim$kG should be apparent in the circular polarization spectrum,
as it is in the case of the magnetic star HD94660.
Fig.\ \ref{fig_observed} shows that a clear Zeeman signature appears in all individual Balmer
lines present in the spectrum of HD94660.
This is evident also from the clear linear relationship between
${\cal F}_V/{\cal F}_I$ and $-C g_{\rm eff} \lambda_0^2 \frac{1}{{\cal F}_I} \frac{{\rm d}{\cal F}_I(\lambda)}{{\rm d}\lambda}$, in contrast to the behavior shown by the central stars of the PNe (Fig.\ \ref{fig_slope}).
Table~1 summarizes the magnetic field inferred from individual lines.
They are all consistent within errors $B_{\rm eff}=-1950\pm 71$~G, as expected from
Landstreet \& Mathys (2000). Moreover, the null spectra ${\cal F}_N/{\cal F}_I$ shows
no spurious polarization effects (Table 1, Fig.\,\ref{fig_slope}).

In order to push forward the detection limit, we decrease the noise level
by adding Balmer lines in the velocity frame.
The line addition technique introduced by Semel \& Li (1996) is a widespread technique
that has been successfully applied to detect Zeeman signatures in a large variety of stars
(e.g., Donati \& Landstreet 2009).
The mean spectral lines thus obtained, of the two central stars of NGC 1360 and LSS 1362,
and the magnetic star HD94660 are represented in  Fig.\,\ref{fig_folded}.
In the selected velocity interval, the r.m.s. of ${\cal F}_V/{\cal F}_I$ and ${\cal F}_N/{\cal F}_I$ spectra are similar,
$\sigma({\cal F}_V/{\cal F}_I) \sim \sigma({\cal F}_N/{\cal F}_I) \sim 1.9\times 10^{-4}$, for NGC 1360 and,
$\sigma({\cal F}_V/{\cal F}_I) \sim \sigma({\cal F}_N/{\cal F}_I) \sim 5.9\times 10^{-4}$, for LSS 1362.
The large difference in the case of HD94660, $\sigma({\cal F}_V/{\cal F}_I) \sim 6 \sigma({\cal F}_N/{\cal F}_I) \sim \times 10^{-4}$,
is a further unquestionable evidence of a strong magnetic field in this star.
Fig.~2 shows how large indeed is the circular polarization in Balmer lines of a star
harboring a $\sim$2~kG field.

At our polarimetric sensitivity (better than previous work by Jordan et al. 2005), we
can say that we do not detect any signal in circular polarization due to the Zeeman effect.
More precisely, the magnetic field is below $\sim$300 G ($\sim$600 G) for the central stars of NGC1360 (LSS1362)
with a probability of 68.2\%, and below $\sim$400 G ($\sim$900 G) with a probability of 95.4\%.
These values correspond to the magnetic field obtained with all the spectral lines plus the
values of the error at 68.2\% and  95.4\% confidence levels, see equation (4).

\section{Conclusions}

Contrary to \cite{Jordan05}, we find no evidence for the existence of kG
magnetic fields in the central stars of the PNe NGC1360 and
LSS1326. Our conclusion is based on spectropolarimetric observations
deeper and at higher spectral resolution than those of
\cite{Jordan05}, as well as on a rigorous analysis of the polarization
signal in several Balmer lines, considered individually or added in
the velocity space.  The upper limits that we found for the
longitudinal magnetic field integrated over all stellar disc is $\sim
300$ and $\sim 600$~G for NGC1360 and LSS1362, respectively.
An application of our method to the \cite{Jordan05} data,
obtained from ESO archive, gives an upper limit of $\sim 400$~G
(NGC1360) and $\sim 600$~G (LSS1362). 

With this conclusion, no evidence is left for magnetic fields on PN
{\bf central stars}.  On the other hand, positive indication of
magnetic fields was obtained for the {\bf nebulae} in a handful of
objects: mG fields were found in the young PN OH~0.9+1.3 by OH
circular polarization \citep{Zijlstra89}, and in the bipolar PNe NGC
7027, NGC 6537, and NGC 6302 by polarimetry of magnetically aligned
dust grains \citep{Greaves02, Sabin07}.  Therefore the negative result
for the two PNe studied in this paper should not stop further efforts
to detect magnetic fields in other PNe central stars. The method
described in this paper, sensitive to $\sim$~kG fields, may be
attempted on other PNe which display morphological features expected
for magnetically active PN central stars, such as elongated bipolar
lobes, jets, and ansae (cf. e.g. \cite{GarciaS99}). It should be
remarked, however, that NGC~1360 was exactly one of these promising
targets, as it possesses polar jets with increasing speed with
distance from the central star, expected for a magnetically collimated
outflow \citep{Garcia08}.  Other morphologies should be tested.

\acknowledgements{Based on observations made with ESO Telescopes
at the Paranal Observatories under programme 386.D-0325(A)
and ESO Science Archive Facility. The Spanish contribution has been funded by the Spanish Ministry of
Science and Innovation under the projects AYA2010-18029 and AYA2007-66804.
}


\begin{thebibliography}{99}
\bibitem[\protect\citeauthoryear{Angel \& Landstreet}{1970}]{Angel70} Angel, J.R.P. \&  Landstreet J.D. 1970, ApJ, 160, L147
\bibitem[\protect\citeauthoryear{Balick \& Frank}{2002}]{BalickFrank2002} Balick, B., \& Frank, A. 2002, ARA\&A, 40, 439
\bibitem[\protect\citeauthoryear{Blackman et al.}{2001}]{Blackman01} Blackman, E.G., Frank, A., Markiel, J.A.,  Thomas, J.H., \& Van Horn, H.M. 2001, Nature, 409, 485
\bibitem[\protect\citeauthoryear{Clough et al.}{2003}]{Clough2005} Clough, S.A., Shephard, M.W.,  Mlawer,  E.J.,  Delamere, J.S. et al. 2005, JQSRT, 91, 233
\bibitem[\protect\citeauthoryear{Corradi}{2006}]{Corradi06} Corradi, R.L.M. 2006, IAU Symposium 234, Cambridge University Press, p.277
\bibitem[\protect\citeauthoryear{Donati \& Landstreet}{2009}]{Donati09} Donati, J-F., \& Landstreet, J.D. 2009, ARA\&A, 47, 333
\bibitem[\protect\citeauthoryear{Garc\'{i}a-Segura et al.}{1999}]{GarciaS99} Garc\'{i}a-Segura, G., Langer, N. R\'ozyczka, M., Franco, J. 1999, ApJ, 517, 767
\bibitem[\protect\citeauthoryear{Garc\'{i}a-D\'{i}az et al.}{2008}]{Garcia08} Garc\'{i}a-D\'{i}az, M.T., L\'{o}pez, J.A., Garc\'{i}a-Segura, G., Richer, M.G.,  Steffen, W. 2008, ApJ, 676, 402
\bibitem[\protect\citeauthoryear{Greaves}{2002}]{Greaves02} Greaves, J.S. 2002, A\&A, 392, L1
\bibitem[\protect\citeauthoryear{Jordan et al.}{2005}]{Jordan05} Jordan, S., Werner, K., \&  O'Toole, S.J. 2005, A\&A, 432, 273
\bibitem[\protect\citeauthoryear{Landstreet}{1982}]{Landstreet82} Landstreet, J. 1982, ApJ 258, 639
\bibitem[\protect\citeauthoryear{Landstreet \& Mathys}{2000}]{Landstreet00} Landstreet, J. \& Mathys G. 2000, A\&A, 359, 213
\bibitem[\protect\citeauthoryear{Leone}{2007}]{Leone07} Leone, F. 2007, MNRAS, 382, 1690
\bibitem[\protect\citeauthoryear{Leone \& Catanzaro}{2004, R = 115\ 000}]{Leone04} Leone, F. \& Catanzaro, G. 2004, A\&A, 425, 271
\bibitem[\protect\citeauthoryear{Leone et al.}{2000}]{Leone00} Leone, F., Catalano, S, \& Catanzaro, G. 2000, A\&A, 355, 315
\bibitem[\protect\citeauthoryear{Leone \& Catanzaro}{2001, R = 15\ 000}]{Leone01} Leone, F. \& Catanzaro, G. 2001, A\&A, 365, 118
\bibitem[\protect\citeauthoryear{Levato et al.}{1996}]{Levato96} Levato, H., Malaroda, S., Morrell, N., Solivella, G. \& Grosso, M. 1997, A\&ASS, 118, 231
\bibitem[\protect\citeauthoryear{Mathys}{1989}]{Mathys89}Mathys, G. 1989, Fundam. Cosmic Phys., 13, 143 
\bibitem[\protect\citeauthoryear{Sabin et al.}{2007}]{Sabin07} 	Sabin, L., Zijlstra, A.A., \&  Greaves, J.S. 2007, MNRAS, 376, 378
\bibitem[\protect\citeauthoryear{Semel \& Li}{1996}]{semel96} Semel, M. \& Li, J. 1996, SoPh, 164, 417
\bibitem[\protect\citeauthoryear{Tsui}{2008}]{Tsui08} Tsui K.H. 2008, A\&A, 491, 671
\bibitem[\protect\citeauthoryear{Zijlstra et al.}{1989}]{Zijlstra89} Zijlstra, A.A., te Lintel Hekkert, P., Pottasch, S.R., Caswell, J.L., Ratag, M., \& Habing, H.J. 1989, A\&A, 217, 157
\end{thebibliography}
\end{document}